\documentclass[acus]{JAC2003}

\usepackage{graphicx}
\usepackage{booktabs}
\usepackage{dblfloatfix}
\usepackage{relsize}
\def\cesrta{{C{\smaller[2]ESR}TA}}

\begin{document}
\title{OBSERVATIONS AND  PREDICTIONS  AT {\cesrta}, AND OUTLOOK FOR ILC\thanks{Work supported by the US National Science Foundation (PHY-0734867, PHY-1002467, and PHY-1068662), US Department of Energy (DE-FC02-08ER41538), and the Japan/US Cooperation Program}}
\author{G. Dugan, M. Billing, K. Butler, J. Chu, J. Crittenden, M. Forster,
D. Kreinick, R. Meller, \\ M. Palmer, G. Ramirez, M. Rendina,
N. Rider, K. Sonnad, H. Williams,\\CLASSE, Cornell University, Ithaca, NY, US\\
R. Campbell, R. Holtzapple, M. Randazzo, Cal. Poly. State University, San Luis Obispo, CA, US\\
J. Flanagan, K.Ohmi, KEK, Tsukuba, Ibaraki, Japan\\
M. Furman, M. Venturini, LBNL, Berkeley, CA, US\\
M.Pivi, SLAC, Menlo Park, CA, US}

\maketitle

\section{introduction}
In this paper, we  will describe some of the recent experimental measurements~\cite{IPAC11, EC10,IPAC12GD} performed at \mbox{{\cesrta}}~\cite{CTAP09}, and the supporting simulations, which probe the interaction of the electron cloud with the stored beam. These experiments have been done over a wide range of beam energies, emittances, bunch currents, and fill patterns, to gather sufficient information to be able to fully characterize the beam-electron-cloud interaction and validate the simulation programs. The range of beam conditions is chosen to be as close as possible to those of the ILC damping ring, so that the validated simulation programs can be used to predict  the performance of these rings with regard to electron-cloud-related phenomena.

Using the new simulation code \texttt{Synrad3D} to simulate the synchrotron radiation environment, a vacuum chamber design has been developed for the ILC damping ring which achieves the required level of photoelectron suppression. To determine the expected electron cloud density in the ring, EC buildup simulations have been done based on the simulated radiation environment and on the expected performance of the ILC damping ring chamber mitigation prescriptions. The expected density has been compared with analytical estimates of  the instability threshold, to verify that the ILC damping ring vacuum chamber design is adequate to suppress the electron cloud single-bunch head-tail instability.
\section{Experimental hardware and techniques}
The principal experimental methods~\cite{PAC11E, IPAC10M, Xray1} used to study the dynamics of the beam in the presence of the electron cloud are:
\begin{itemize}
\item bunch-by-bunch tune measurements using one or more gated BPM's, in which a whole train of bunches is coherently excited, or in which individual bunches are excited;
\item bunch-by-bunch frequency spectral measurements of self-excited bunch trains, using a high-sensitivity, filtered and gated BPM, and a spectrum analyzer;
\item bunch-by-bunch, turn-by-turn beam size measurements of self-excited bunch trains, using an x-ray beam size monitor~\cite{Xray1}; and
\item damping rate measurements of individual bunches in trains excited in dipole and head-tail modes, using a high-sensitivity, filtered and gated BPM, a spectrum analyzer, a transverse kicker and an RF-cavity phase modulator.
\end{itemize}
The last technique is still being developed, and will not be discussed further in this paper. It can potentially be very useful, since measurements of bunch-by-bunch damping rates provide additional information on the nature of the effective impedance of the cloud which cannot be obtained in any other way.

\section{Coherent tune measurements}
 \begin{figure*}[!htb]
  \centering$
\begin{array}{cc}
  \includegraphics[width=3.3in]{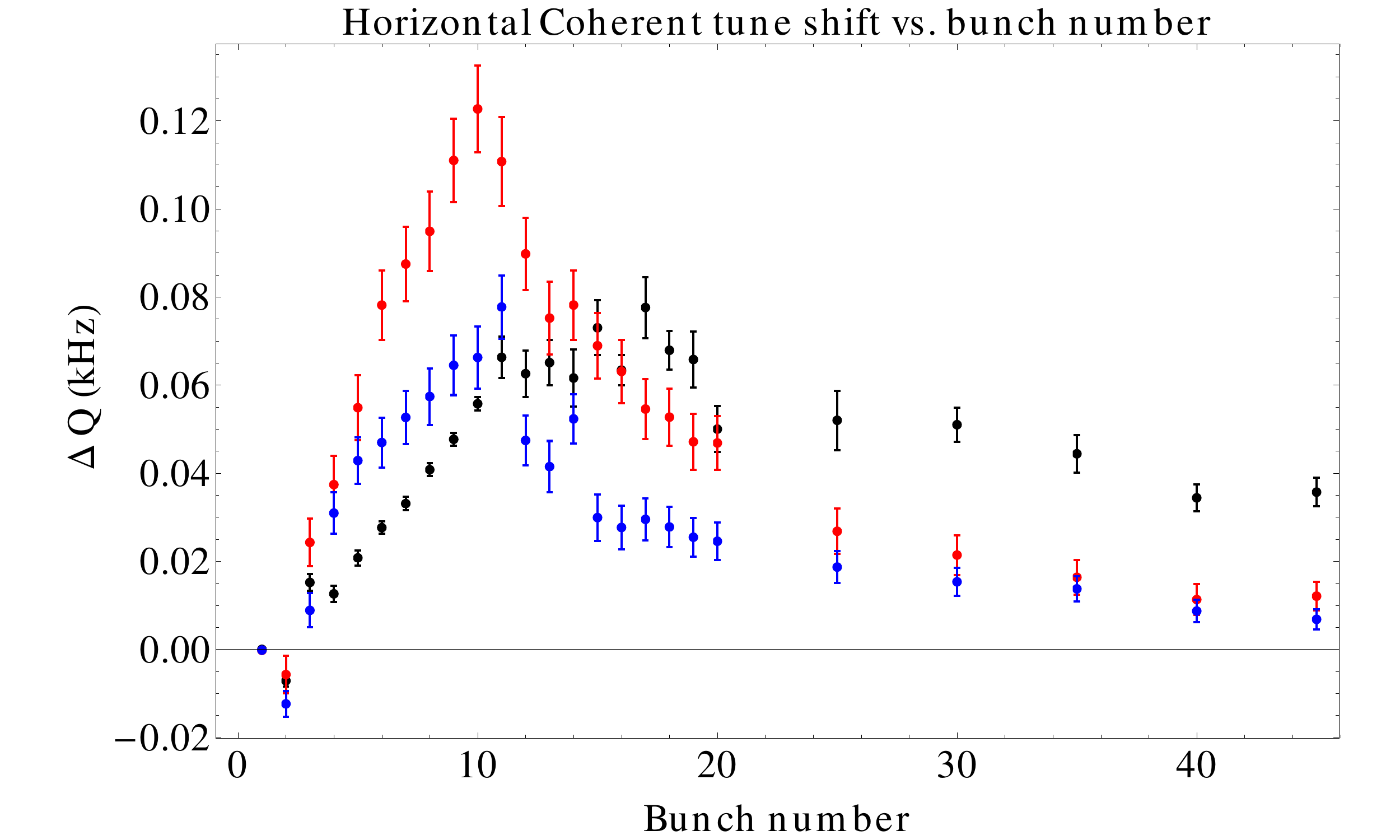}&
    \includegraphics[width=3.3in]{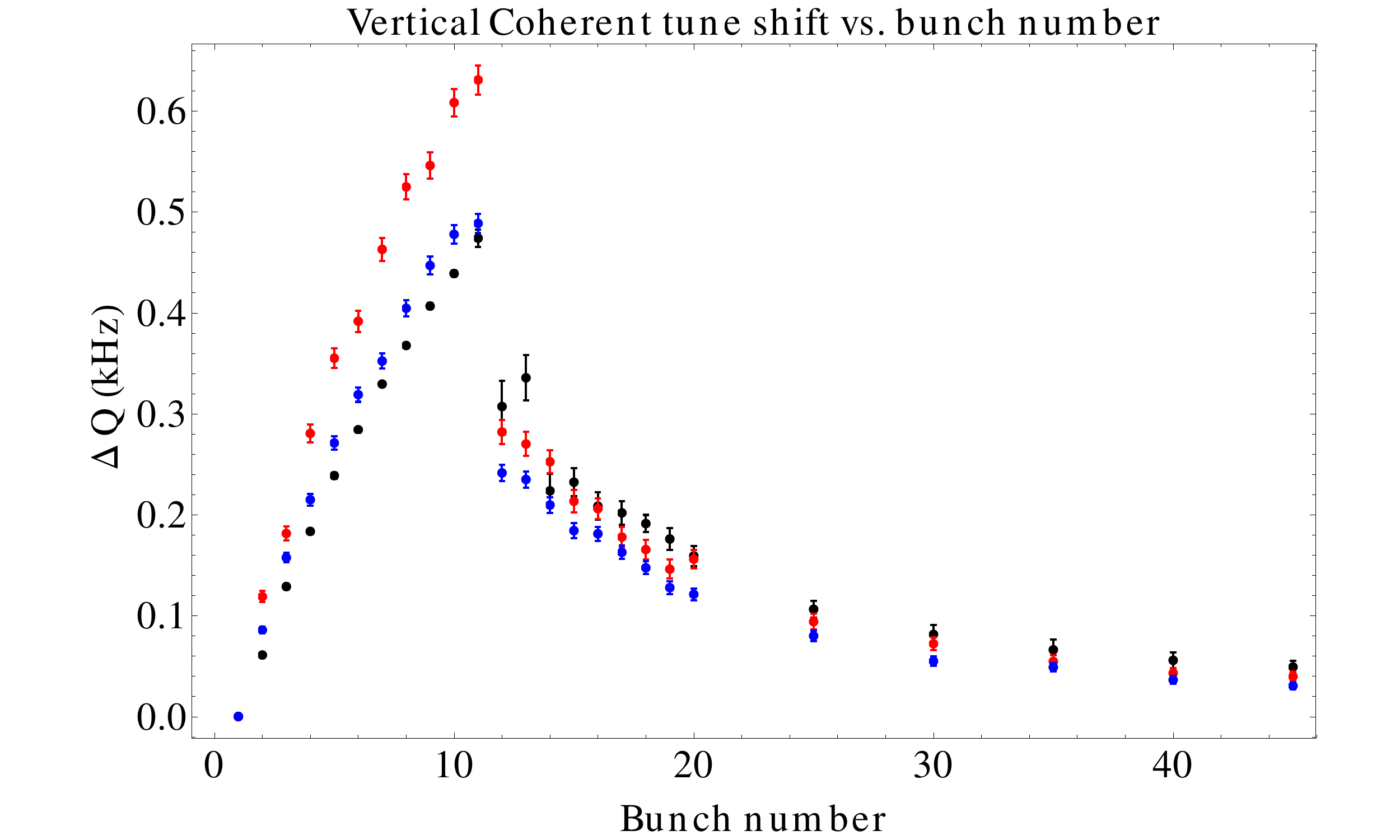}
   \end{array}$
   \caption{
                \label{fig:HV_s2d_s3d_comp-53}
                Measured tune shifts (black points) vs. bunch number, for a train of 10 0.75~mA/bunch 5.3~GeV positron bunches with 14 ns spacing, followed by witness bunches at various spacings. Left: horizontal, on an expanded scale. Right, vertical. Red points are computed (using \texttt{POSINST}) based on direct radiation and an ad-hoc assumption about the scattered photons. Blue points, which are in better agreement with the data, are computed using results from \texttt{Synrad3D} as input to \texttt{POSINST}.  }
   \end{figure*}
A large variety of coherent tune shift data have been taken, covering a wide range of beam and machine conditions.
The contribution to the bunch-by-bunch tune shifts from drift and dipole beamline elements have been computed from the electric field gradients of the charge distributions predicted by the electron cloud simulation codes. The ringwide average tune shifts were then calculated by taking a beta-weighted average of the field gradients per beamline element, and compared with measurements.

\begin{figure*}[!hb]
  \centering
  \includegraphics[width=6.65in]{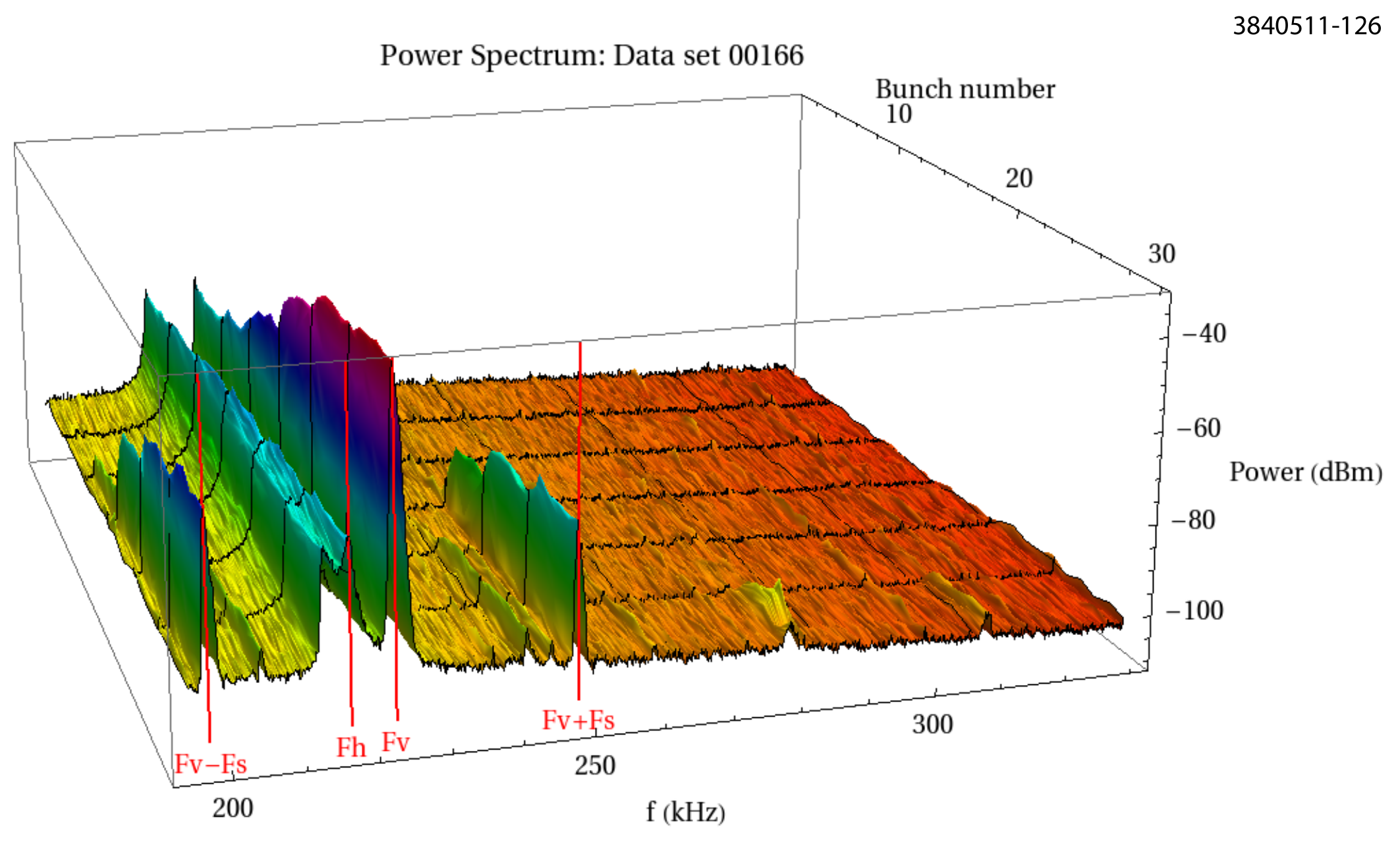}
   \caption{
    \label{fig:spectral_plot_00166}
    Bunch-by-bunch power spectrum for a 30 bunch train of 0.75 mA/bunch positrons at 2 GeV, with a bunch spacing of 14 ns.}
   \end{figure*}

Quite good agreement~\cite{PAC11K,EC10K, IPAC10C, PAC09} has been found between the measurements and the computed tune shifts,
 using either of the buildup codes \texttt{POSINST}~\cite{POSINST-ref}  or \texttt{ECLOUD}~\cite{ECLOUD-ref}. This agreement, which is found for the same set of simulation parameters applied across a wide variety of machine conditions, both constrains many of the model parameters and gives confidence that the models do in fact predict accurately the average density of the electron cloud measured in \mbox{{\cesrta}}.

To help characterize the photoelectrons which seed the cloud in \mbox{{\cesrta}}, and to allow accurate extrapolation to other radiation environments, a new simulation program, \texttt{Synrad3D}~\cite{S3D,S3D2}, has been developed, which predicts the distribution and energy of absorbed synchrotron radiation photons around the ring, including specular and diffuse scattering in three dimensions, using a detailed model for the vacuum chamber geometry. The output from this program can be used as input to the cloud buildup codes, thereby eliminating the need for any ad-hoc assumptions about the photon distributions. Tune shifts computed from buildup simulations with input from \texttt{Synrad3D} agree well with measurements (see Fig.~\ref{fig:HV_s2d_s3d_comp-53}).
   \begin{figure*}[!ht]
  \centering$
\begin{array}{cc}
\includegraphics[width=3.2in]{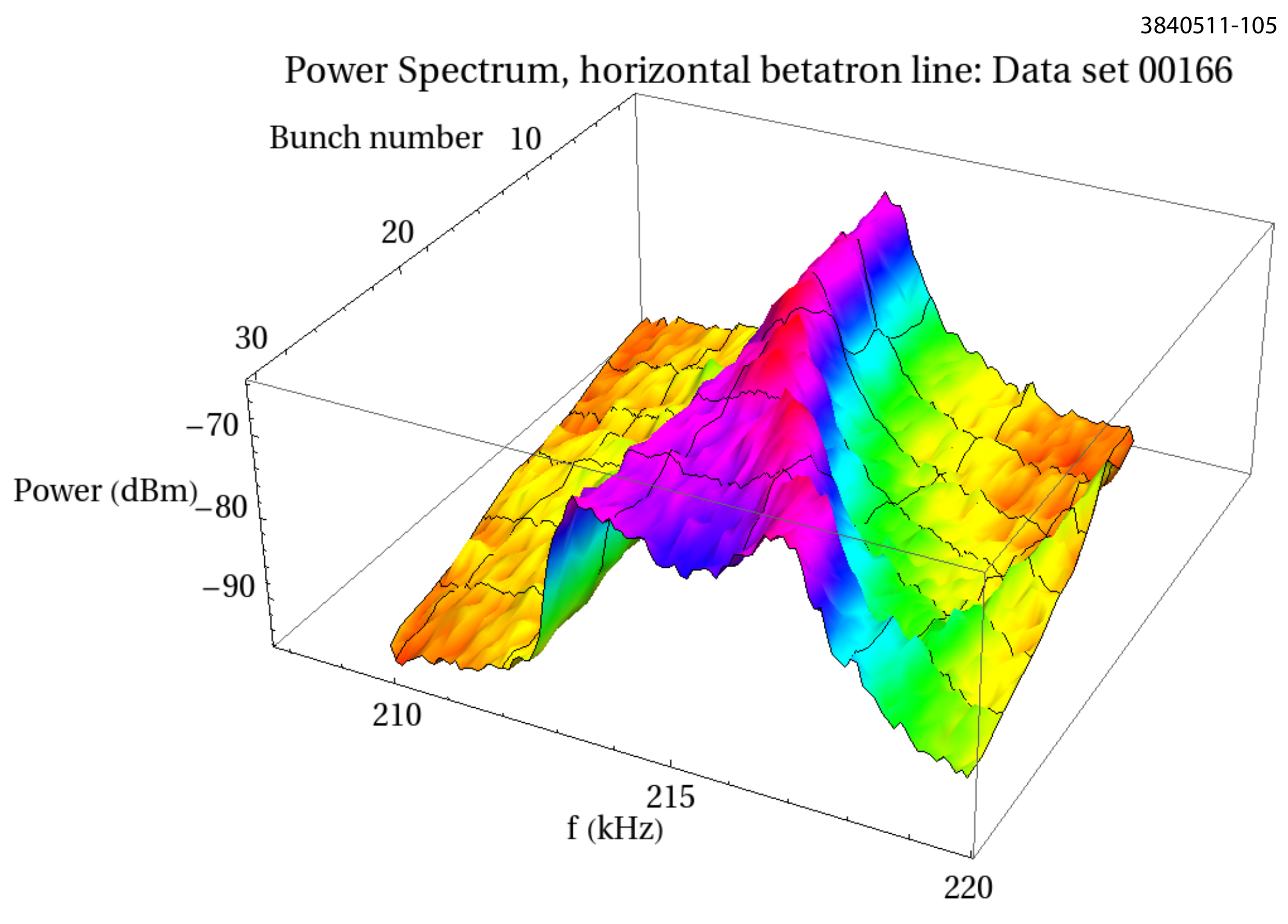}&
  \includegraphics[width=3.2in]{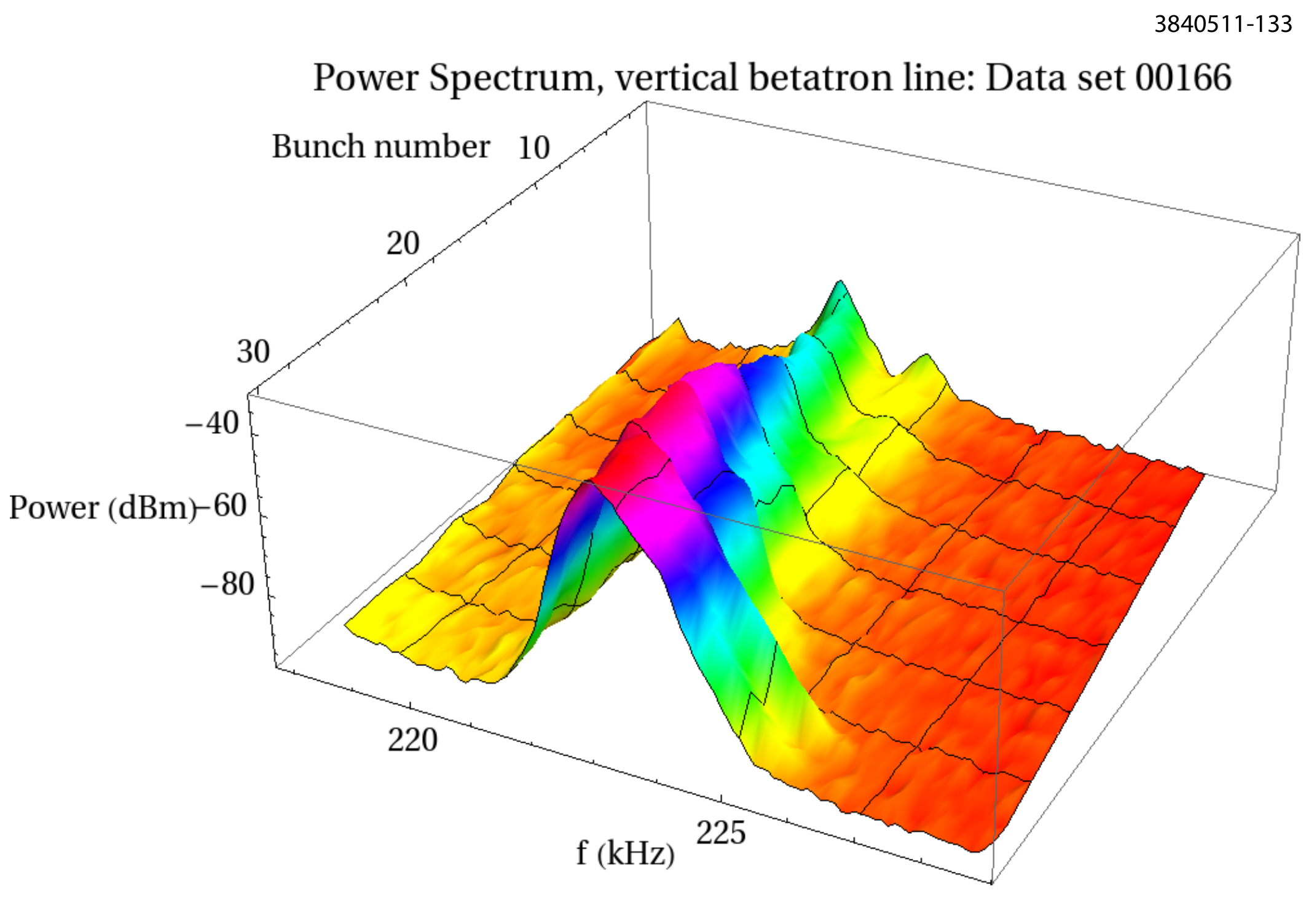}
     \end{array}$
      \caption{\label{fig:HV_spectral_plot_00166}Bunch-by-bunch power spectrum: Left, detail at Horizontal betatron line. Right, detail at vertical betatron line.}
   \end{figure*}
   \begin{figure*}[!hb]
  \centering$
  \begin{array}{cc}
  \includegraphics[width=3.2in]{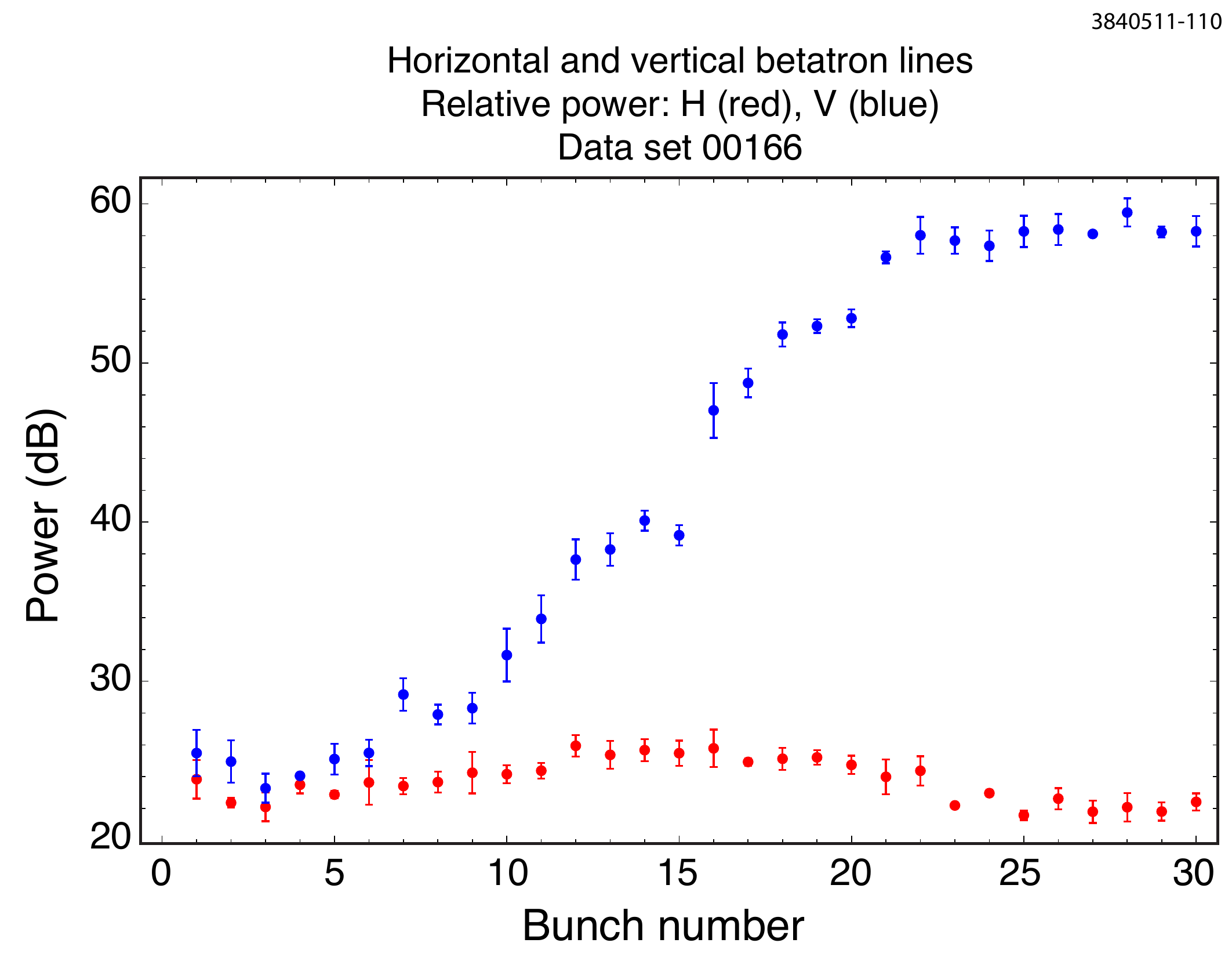}&
    \includegraphics[width=3.2in]{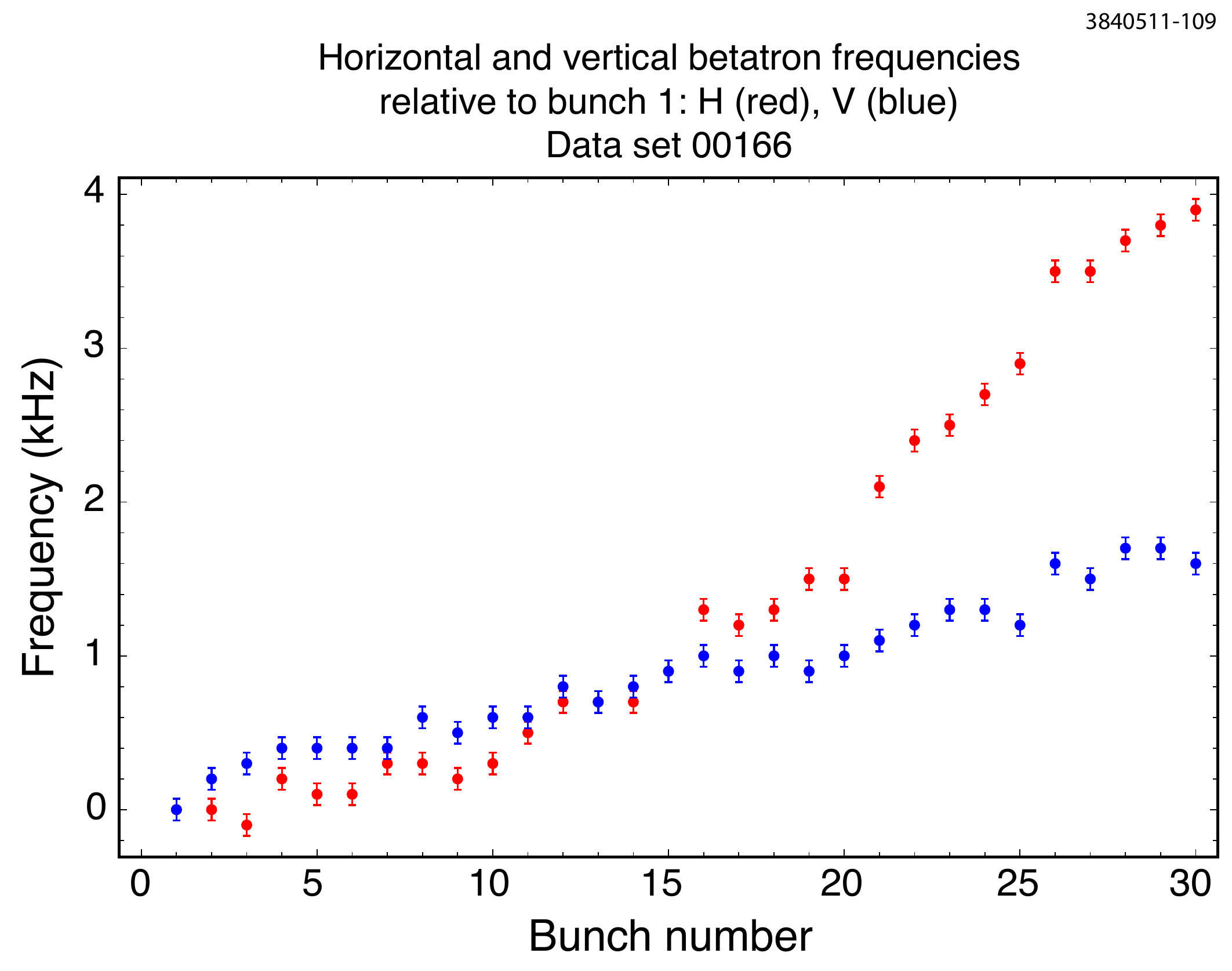}
    \end{array}$
   \caption[Left, Horizontal and vertical peak power  and frequency vs. bunch number]{
                \label{fig:HV_power_freq_plot_00166}
               Left, horizontal and vertical peak power vs. bunch number; right,  horizontal and vertical tune shifts  vs. bunch number.}
   \end{figure*}
\section{Instability threshold measurements}
\label{instab_th}

   Using a high-sensitivity, filtered and gated BPM, and a spectrum analyzer, bunch-by-bunch frequency spectra have been collected for a variety of machine and beam conditions, to detect signals of single-bunch instabilities which develop along trains of positron bunches. The BPM button is located at about  45$^{\circ}$ above the bend plane, so it is roughly equally sensitive to horizontal and vertical beam motion.
Under conditions in which the beam is self-excited via the electron cloud, these frequency spectra exhibit  vertical $m = \pm1$ head-tail (HT) lines, separated from the vertical betatron line by  approximately the synchrotron frequency, for many of the bunches along the train. The amplitude of these lines typically (but not always) grows along the train. 

\begin{figure}[ht!]
  \centering
  \includegraphics[width=3.2in]{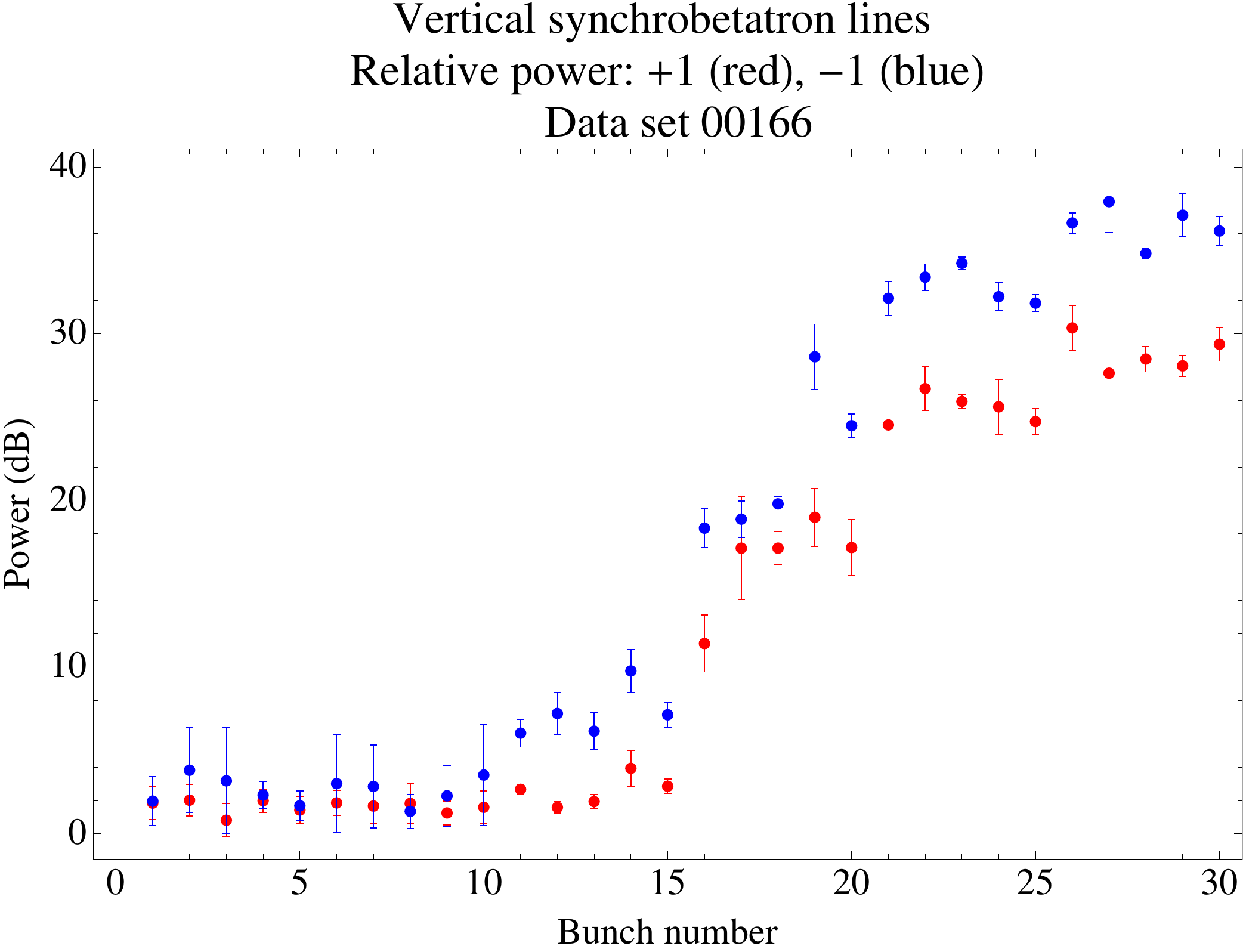} 
   \caption{ \label{fig:fslines_pow_freq-1}
   Vertical head tail lines, peak power vs. bunch number. 2.1 GeV beam energy. Bunch current = 0.75~mA.}
   \end{figure}

An example of a bunch-by-bunch power spectrum illustrating these features is presented in Fig.~\ref{fig:spectral_plot_00166}, which shows the spectral power vs. frequency for each of the bunches along the train. This figure, together with Fig.~\ref{fig:HV_spectral_plot_00166}, Fig.~\ref{fig:HV_power_freq_plot_00166}, Fig.~\ref{fig:EC_BeamDynamics:Cloud_density_comp}, Fig.~\ref{fig:fslines_pow_freq-1}, and Fig.~\ref{fig:Vfspower_plot_00154_00166} (red points), all correspond to the same data set. For this data set, a bunch train of 30 bunches, with an average bunch current of 0.74~mA and a bunch spacing of 14 ns, was observed. The beam energy was 2.1 GeV. 
The horizontal and vertical betatron lines, whose frequencies are labelled by $F_{h}$ and $F_{v}$, respectively, are prominent for all the bunches in the train. The vertical betatron line grows along the train. The vertical $m = \pm1$ head-tail (HT) lines, separated from the vertical betatron line by  approximately the synchrotron frequency, emerge from the noise floor around bunch 15, and also grow in amplitude for later bunches. 
      
     \subsection{Horizontal and vertical betatron lines}
   Fig.~\ref{fig:HV_spectral_plot_00166}, left, shows the bunch-by-bunch power spectrum near the horizontal betatron line.  There is a major peak which shifts up in frequency by about 4~kHz during the bunch train. This shift is attributable to the electron cloud. In addition, there is a lower amplitude ``shoulder'', which appears to be roughly constant in frequency during the bunch train. This bifurcation of the horizontal line is attributable to the dependence of the horizontal tune shift on the mode of multi-bunch oscillation of the train. Similar observations have been made at the SPS~\cite{Corn} and were analyzed using a simple matrix model to describe the beam-cloud interaction in dipoles.

  \begin{figure}[ht!]
  \centering
  \includegraphics[width=3.2in]{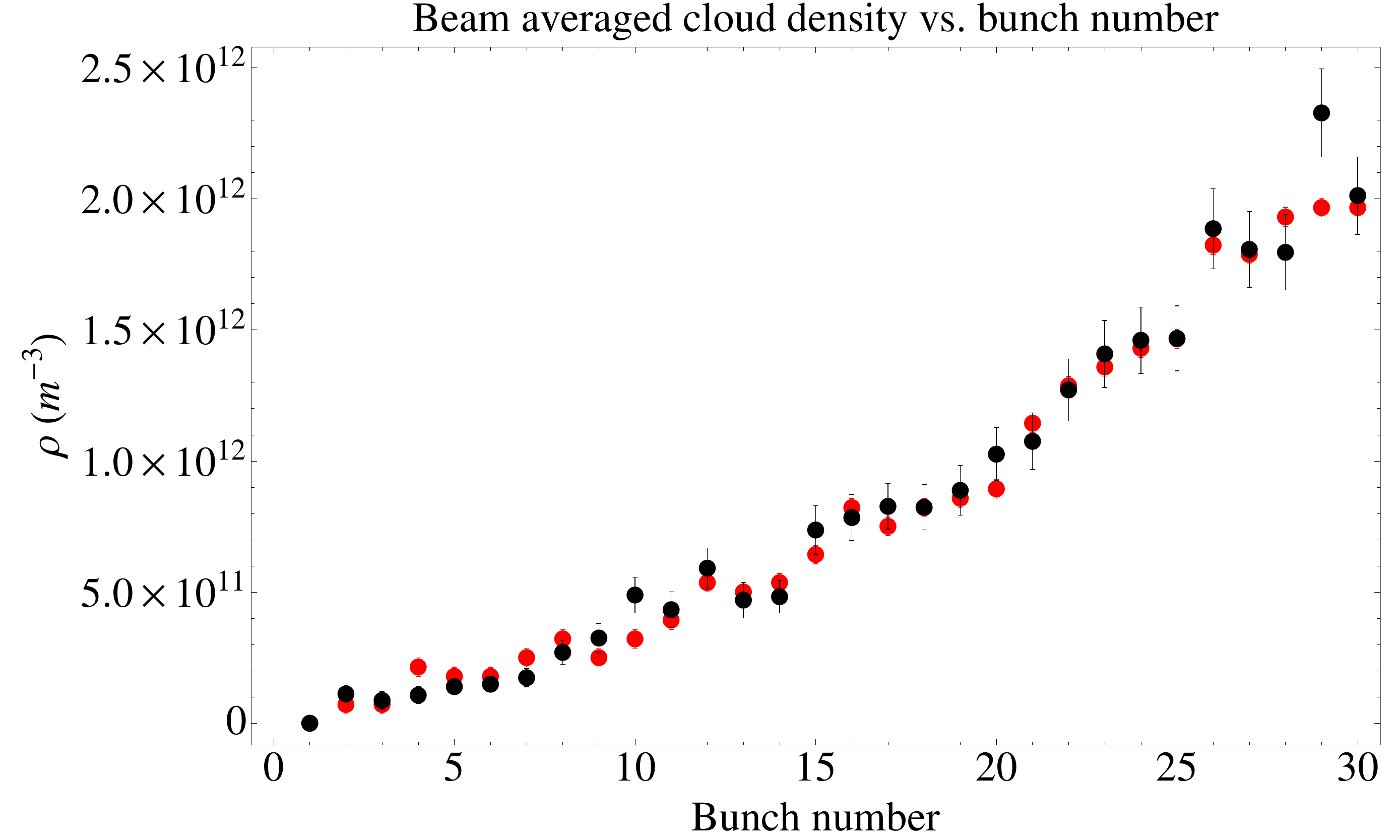}
   \caption[Average initial (i.e., before the ``pinch'') electron cloud density vs. bunch number]{\label{fig:EC_BeamDynamics:Cloud_density_comp}
   Average initial (i.e., before the ``pinch'') electron cloud density vs. bunch number, comparison between estimate from measured tune shifts (red), and simulation (black)  from \texttt{POSINST}.}
   \end{figure}

     \begin{figure}[ht!]
  \centering
   \includegraphics[width=3.2in]{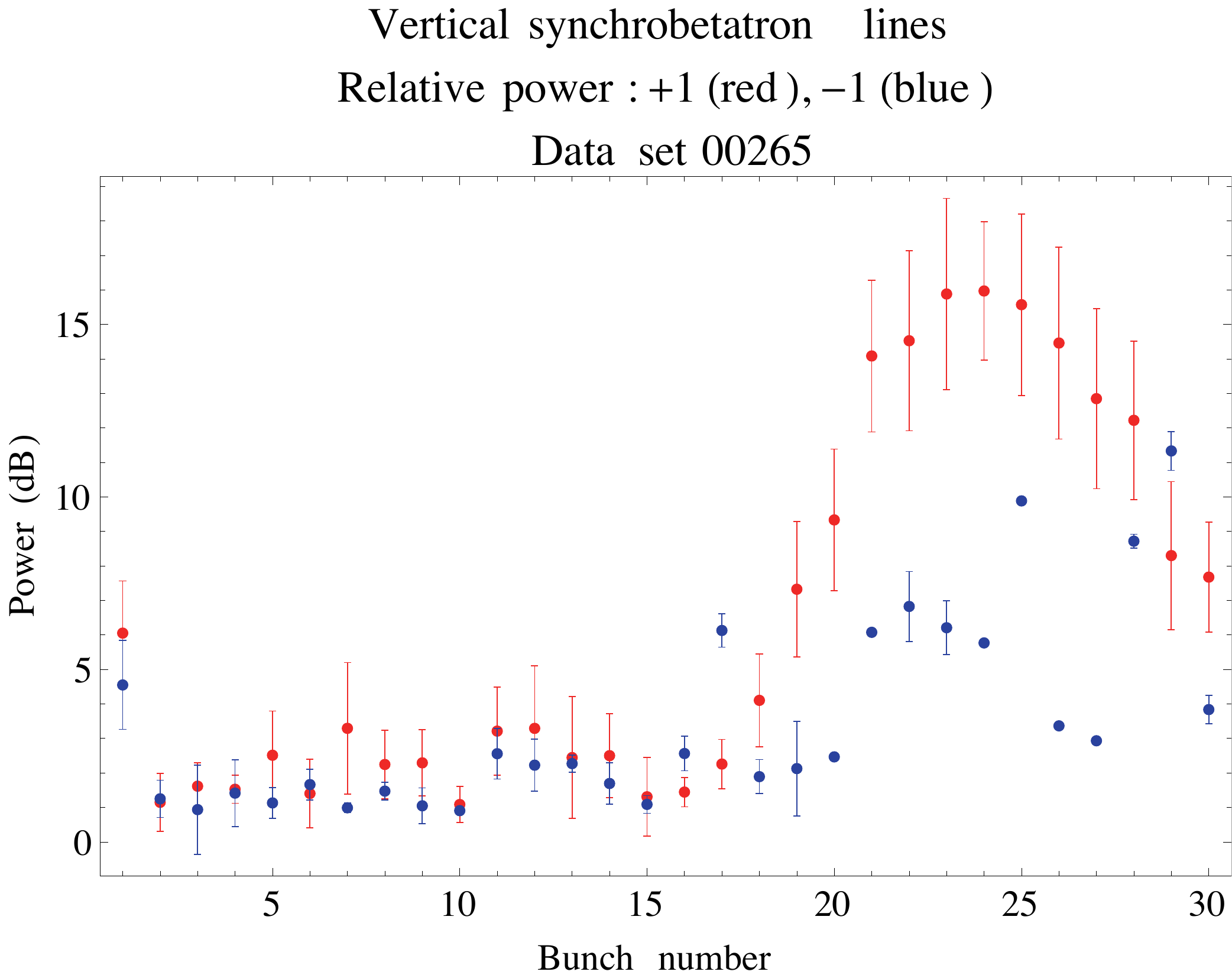}
 \caption{\label{fig:fslines_pow_freq-2}
   Vertical head tail lines, peak power vs. bunch number. 4 GeV beam energy. Bunch current = 1.1~mA. }
   \end{figure}

   Fig.~\ref{fig:HV_spectral_plot_00166}, right, shows the bunch-by-bunch power spectrum near the vertical betatron line. In this case, there is a shift up in frequency of the major peak by about 2~kHz during the bunch train, which is again attributable to the electron cloud. In addition, there is a smaller peak at a higher frequency, present even on the first bunch, which appears to grow in amplitude and merge with the main peak near the end of the bunch train. This structure in the vertical line is not understood, but is likely to be a single-bunch effect.
   
   The peak power in the horizontal and vertical lines, and the tune shift of these lines along the train, is plotted vs bunch number in  Fig.~\ref{fig:HV_power_freq_plot_00166}.  The electron cloud density near the beam can be determined directly from the tune shifts using the approximate relation
         \begin{equation}
         \left<\rho_{c}\right>=\gamma\frac{\Delta Q_{x}+\Delta Q_{y}}{r_{e}\left<\beta\right> C},
         \end{equation}
       in which $\left<\beta\right>$ is the average beta function, $C$ is the ring circumference, $\gamma$ is the beam Lorentz factor, and $r_{e}$ is the classical electron radius.  Alternatively,  the corresponding density can be obtained from a simulation which is adjusted to predict the measured tune shifts shown in the right plot in Fig.~\ref{fig:HV_power_freq_plot_00166}.  The cloud density vs. bunch number extracted from the tune shifts is shown in Fig.~\ref{fig:EC_BeamDynamics:Cloud_density_comp}.

 %               \FloatBarrier

   \subsection{Head-tail lines and instability threshold determination}
We attribute the presence of the head-tail lines in Fig. ~\ref{fig:spectral_plot_00166} to a vertical head-tail instability induced by the electron cloud. In Fig.~\ref{fig:fslines_pow_freq-1}, the peak power in these lines is plotted vs. bunch number.  
 By comparing this figure with the cloud density-bunch number correlation shown in Fig.~\ref{fig:EC_BeamDynamics:Cloud_density_comp}, we can conclude that the onset of the HT lines occurs at a ringwide initial (i.e., before the ``pinch'') beam-averaged cloud density  of around $8\times 10^{11}$~m$^{-3}$ for 2.1 GeV beam energy. 
 
 Using this same method, the head-tail line data shown in Fig.~\ref{fig:fslines_pow_freq-2} were used to establish that the corresponding threshold density at 4 GeV is about $2\times 10^{12}$~m$^{-3}$.
       
  \begin{figure*}[ht!]
  \centering
  \includegraphics[width=6.6in]{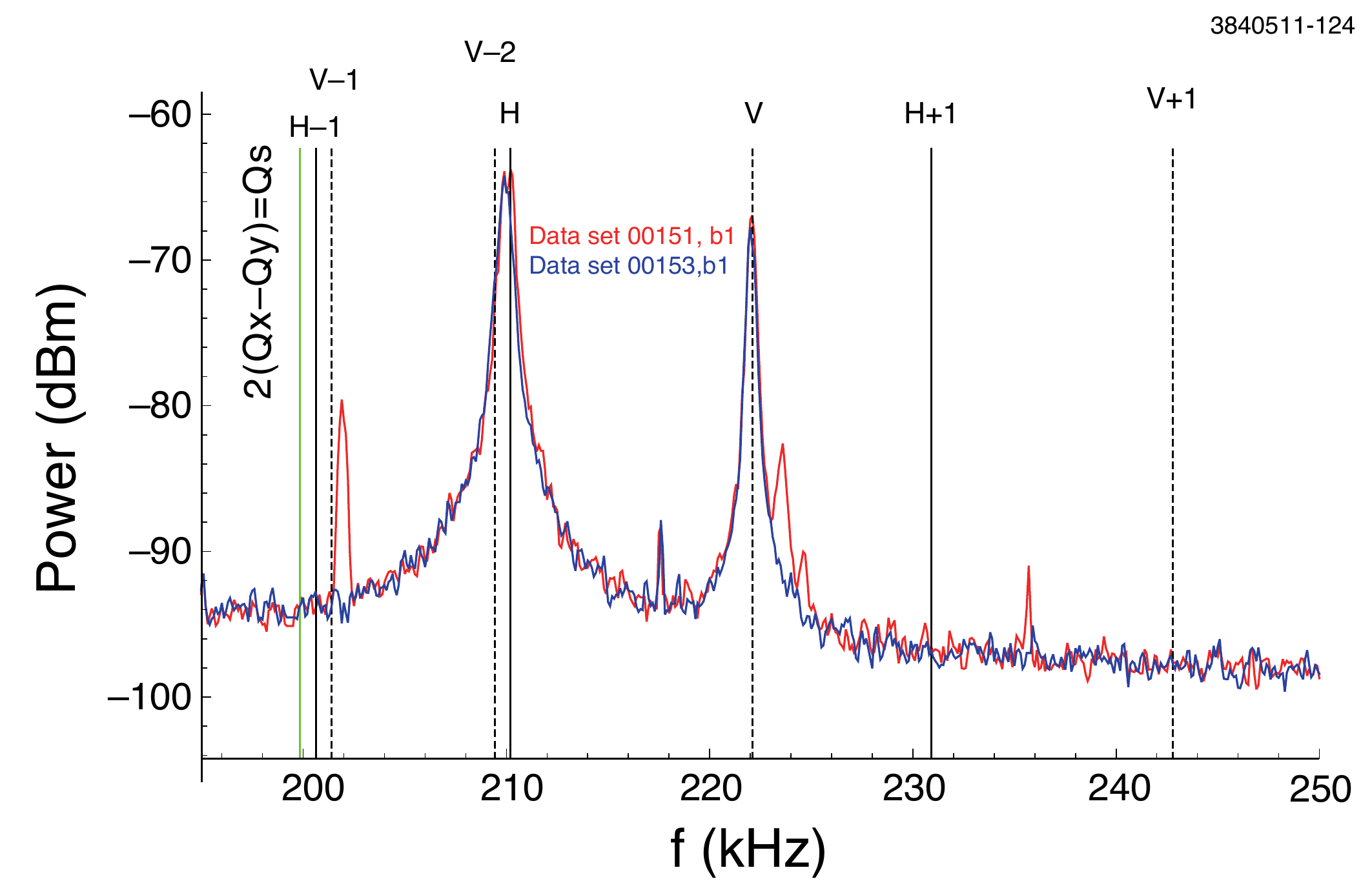}
   \caption[Data set 151 and 153 : Power spectrum, bunch 1 compared.]{ \label{fig:Spectral_plot_00151_1_00153_1}
   Data set 151 and 153 : Power spectrum, bunch 1, compared. Both sets have the same chromaticity and bunch current, but there is a precursor bunch present for data set 153, as described in the text. The lines labelled, for example,``V+1''  and ``V-1'' are shown at frequencies of $\pm f_{s}$ from the vertical betatron line (``V''), in which $f_{s}$ is the synchrotron frequency. For these data sets, $f_{s}~=~20.7$~kHz. The location of a machine resonance is also indicated.}
   \label{f27}
   \end{figure*}

    \begin{figure*}[ht!]
\centering
  \includegraphics[width=6.6in]{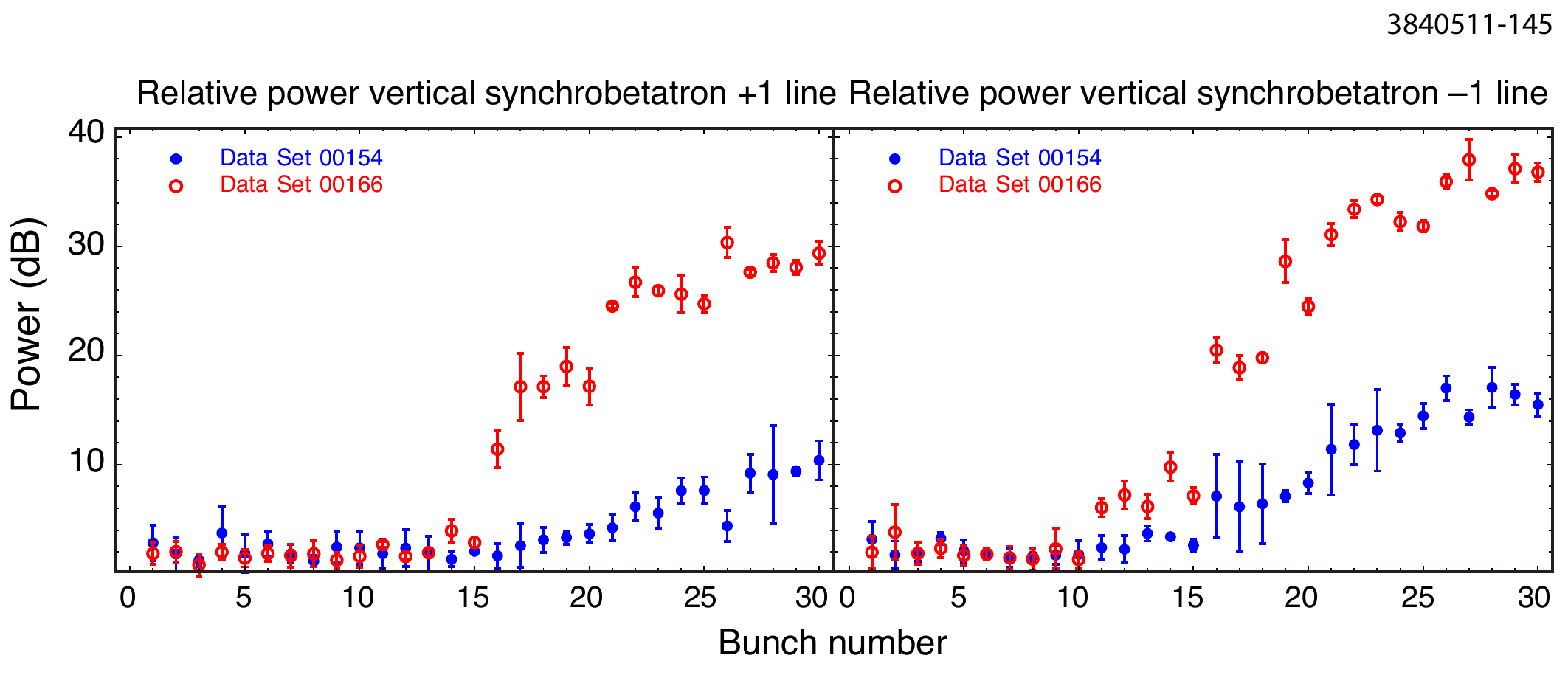}
   \caption[Vertical head-tail lines compared.]{\label{fig:Vfspower_plot_00154_00166}
   Vertical head-tail lines for electron and positron beams. Both data sets have the same chromaticity and bunch current, but data set 166 was taken with a positron beam while data set 154 was taken with an electron beam.}
   \end{figure*} 

         \subsection{Instability observations for the first bunch in the train.}
        
 Under some conditions, the first bunch in the train also exhibits head-tail lines.

     In Fig.~\ref{fig:Spectral_plot_00151_1_00153_1}, the power spectrum of bunch~1 for data set 151 is shown (red trace). Note the presence of a prominent $m=-1$ head-tail line. This line disappears for the second bunch, and does not re-appear until much later in the train (see Fig.~\ref{fig:fslines_pow_freq-2}). Moreover, beam size measurements (see Fig.~\ref{fig:welh08_fig235} (center and bottom plots), below) indicate that the first bunch in the train is frequently larger in size than the next few bunches. 
              
              One possible explanation for this observation is that there is trapped cloud density near the beam, which persists long after the train ends, that is sufficiently high, even for the first bunch in the train, to induce spontaneous head-tail motion.  However, the interaction of the first bunch with this trapped cloud evidently destabilizes it, causing it to disperse, so that bunch 2 does not suffer from spontaneous head-tail motion. 
              
              Simulations and witness bunch tune measurements indicate that the electron cloud lifetime in dipoles and drifts is much shorter than one turn in {\cesrta}. Cloud density which persists for many turns may be due to trapped cloud in quadrupoles and wigglers. RFA measurements and simulations indicate this trapped cloud may be in the quadrupoles~\cite{LW} and wigglers~\cite{CC}.
              
                 To test this hypothesis, in data set 153, a 0.75~mA ``precursor'' bunch was placed 182 ns before bunch 1. Otherwise, conditions were the same as for data set 151. The spectrum of the first bunch for data set 153 is shown (blue trace) in Fig.~\ref{fig:Spectral_plot_00151_1_00153_1}. Note that the lower head-tail line is now absent. In addition, the structure seen on the upper edge of the vertical betatron line in Fig.~\ref{fig:Spectral_plot_00151_1_00153_1} is disappears.                     \subsection{Other observations}
    Other observations from systematic studies are:
   %%%%%%%%%%%%%
  \begin{itemize}
\item The onset of the HT lines depends strongly on the vertical chromaticity, the beam current, the number of bunches, and the bunch spacing.
\item
For a 45~bunch train, the HT lines have a maximum power  around bunch~30; the line power is reduced for later bunches.
\item There is a weak dependence of the onset of the HT lines on the synchrotron tune, the single-bunch vertical emittance, and the vertical feedback.
\item Under identical conditions, HT lines also appear in electron trains, but the onset is later in the train, develops more slowly, and is much weaker, than for positrons.
This is illustrated in Fig.~\ref{fig:Vfspower_plot_00154_00166}.

\item The HT line structure observed for the last bunch in a 30~bunch train varies strongly as a function of the current in that bunch.  But the frequency of the vertical betatron line of this bunch is  only very weakly dependent on the current in the bunch.
\end {itemize}

\section{Measurements of emittance growth along bunch trains}
Using an x-ray monitor~\cite{Xray1}, bunch-by-bunch beam position and size measurements have been made on a turn-by-turn basis for positron beams. From the beam size measurements, the evolution of the beam emittance along trains of bunches has been measured. 

\begin{figure}[htb!]
   \centering
 \includegraphics[width=3.2in]{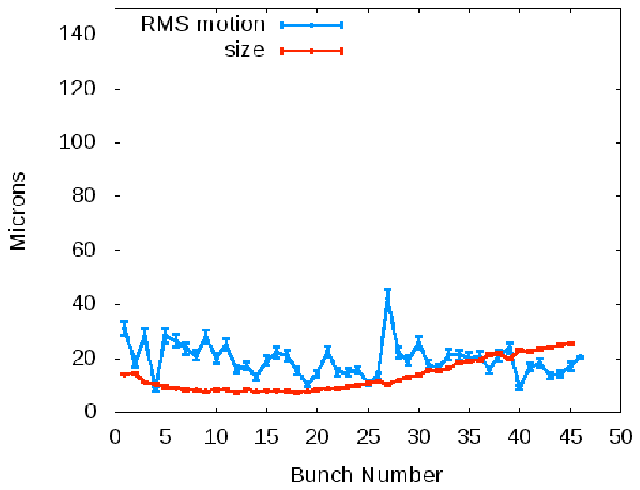}
    \includegraphics[width=3.2in]{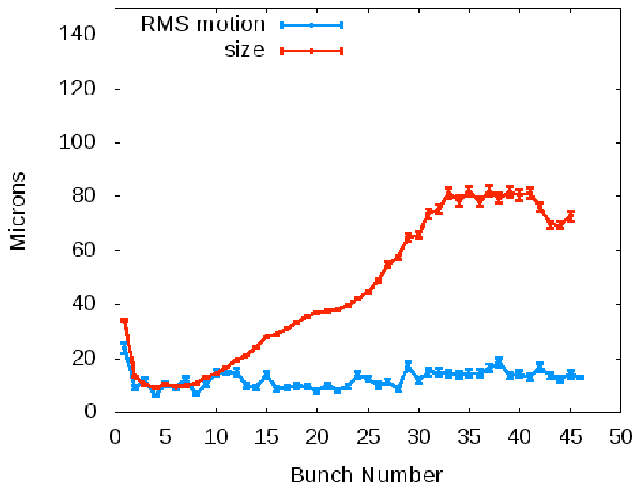}
        \includegraphics[width=3.2in]{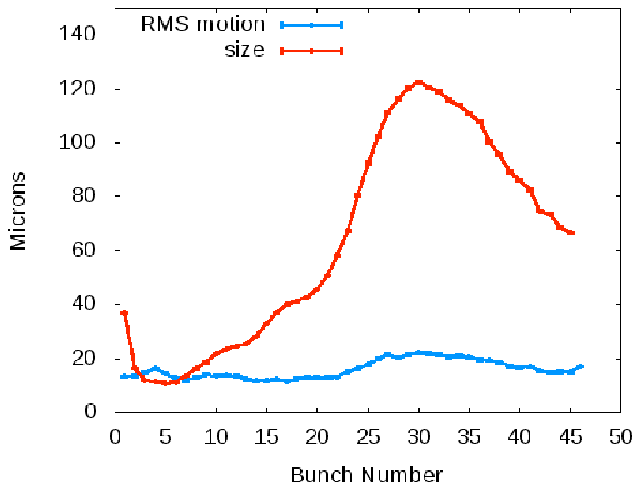}
   \caption{\label{fig:welh08_fig235}Bunch-by-bunch beam size and rms motion at 14 ns spacing and 2.1 GeV. Top: bunch current 0.5 mA/bunch. Center: bunch current 1 mA/bunch. Bottom: bunch current 1.3 mA/bunch.}
\end{figure}

Beam centroid motion and vertical emittance are observed to grow along the train. The growth pattern is a strong function of the bunch current (see Fig.~\ref{fig:welh08_fig235}). Often, the first bunch in the train has an anomalously large size, which correlates with of the observation of a vertical head-tail line in the spectrum of this bunch, as discussed above.

   \begin{figure}[ht!]
   \centering
    \includegraphics[width=3.2in]{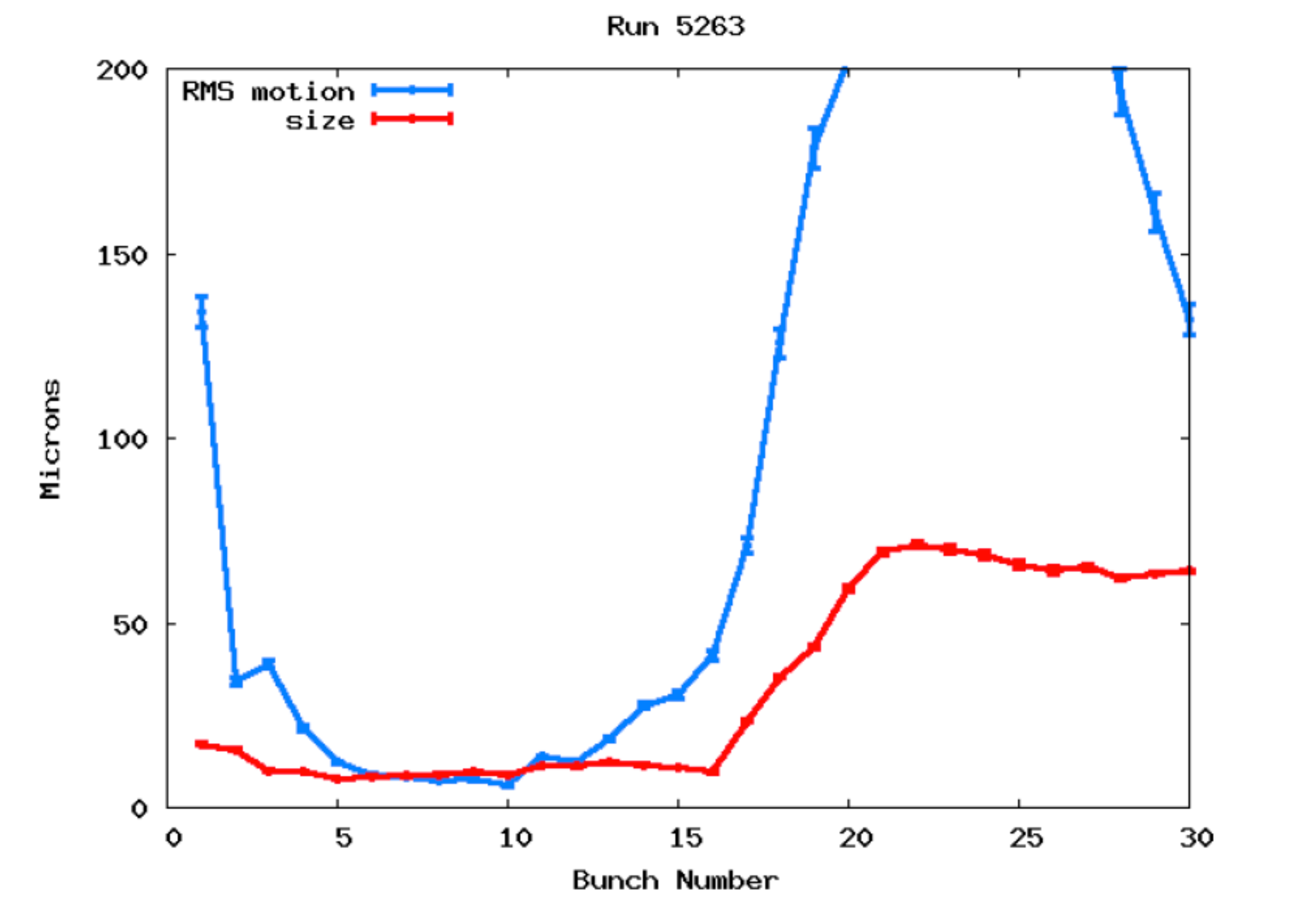}
   \caption{\label{fig:run_5263}Bunch-by-bunch beam size and rms motion at 14~ns, 4 GeV, with 1.1 mA/bunch.}
\end{figure}

\begin{table*}[hb!]
\caption{\label{tab:thr}Analytical estimates of EC instability thresholds}
\begin{center}
\begin{tabular}{lccc}
\hline
& {\cesrta} 2.1~GeV & {\cesrta} 4~GeV &ILC DR (DTC04)  \\
\hline
Circumference $C$ (m)  &  768 & 768  &3278\\
Energy $E$ (GeV)  & 2.1 & 4.0  &5.0 \\
Bunch population $N_{+}$ ($\times10^{10}$) & 2 & 2  &2\\
Emittance ($\varepsilon_x$ (nm), $\varepsilon_y$ (pm) )& (2.6, 20) & (40, 177)  & (0.57, 2)\\
Momentum compaction $\alpha$ ($\times10^{-4}$) &68.0 & 63.0 &3.3 \\
Rms bunch length $\sigma_z$ (mm) & 10.5 &  17.2  &6\\
Rms energy spread $\sigma_E/E$ ($\times10^{-3}$) & 0.81 & 0.93 &1.09\\
Horizontal betatron tune $\nu_x$ & 14.57  & 14.57  &47.37 \\
Vertical betatron tune $\nu_y$ &9.62  &9.62   &28.18 \\
Synchrotron tune $\nu_s$ & 0.065 & 0.041  &0.031\\
Damping time $\tau_{x,y}$ (ms)  & 56.4 & 19.5 & 23.8\\
Average vertical beta function $\left<\beta_{y}\right>$ (m)  & 16 & 16&24\\
Electron frequency $\omega_e/2\pi$ (GHz)  & 43 & 11.3 &105.8  \\
Phase angle $\chi$  & 9.5 & 4.1 &13.3\\
Threshold density $\rho_{e,th}$ ($\times10^{12}$~m$^{-3}$) & 1.3 & 2.65 &0.23\\
Tune shift at threshold  $\Delta\nu_{x+y}$ ($\times10^{-3}$)& 10.7 & 11.6 &5.1\\
\hline
\end{tabular}
\end{center}
\end{table*}

In Fig.~\ref{fig:run_5263}, the bunch-by-bunch beam size and rms motion are shown for a measurement with a 14~ns train, at 4 GeV, with 1.1 mA/bunch. The conditions for this measurement are exactly the same as those for the bunch-by-bunch frequency measurement whose head-tail line growth is shown in Fig.~\ref{fig:fslines_pow_freq-2}.  The $m=1$ vertical head-tail line starts growing at bunch~18 and peaks around bunch~22. Comparing with Fig.~\ref{fig:run_5263}, the vertical emittance growth starts at bunch 17 and reaches a plateau around bunch~22.  Thus the onset and development along the train of the vertical head-tail line is very similar to the onset  and development along the train of  vertical emittance growth. 

Other key observations are:
\begin{itemize}
\item The threshold for beam size growth along the train is not very sensitive to the chromaticity or the bunch spacing, although the maximum beam size along the train is  larger for a smaller chromaticity. This dependence on the chromaticity is in contrast to the behavior of the head-tail lines, which are quite sensitive to chromaticity. 
\item Beam size growth along the train is also not very sensitive to the initial beam size or the feedback gain.
\end{itemize}

\section{Comparisons with analytic estimates and simulations}

 \begin{figure*}[ht!]
  \centering$
\begin{array}{cc}
  \includegraphics[width=3.3in]{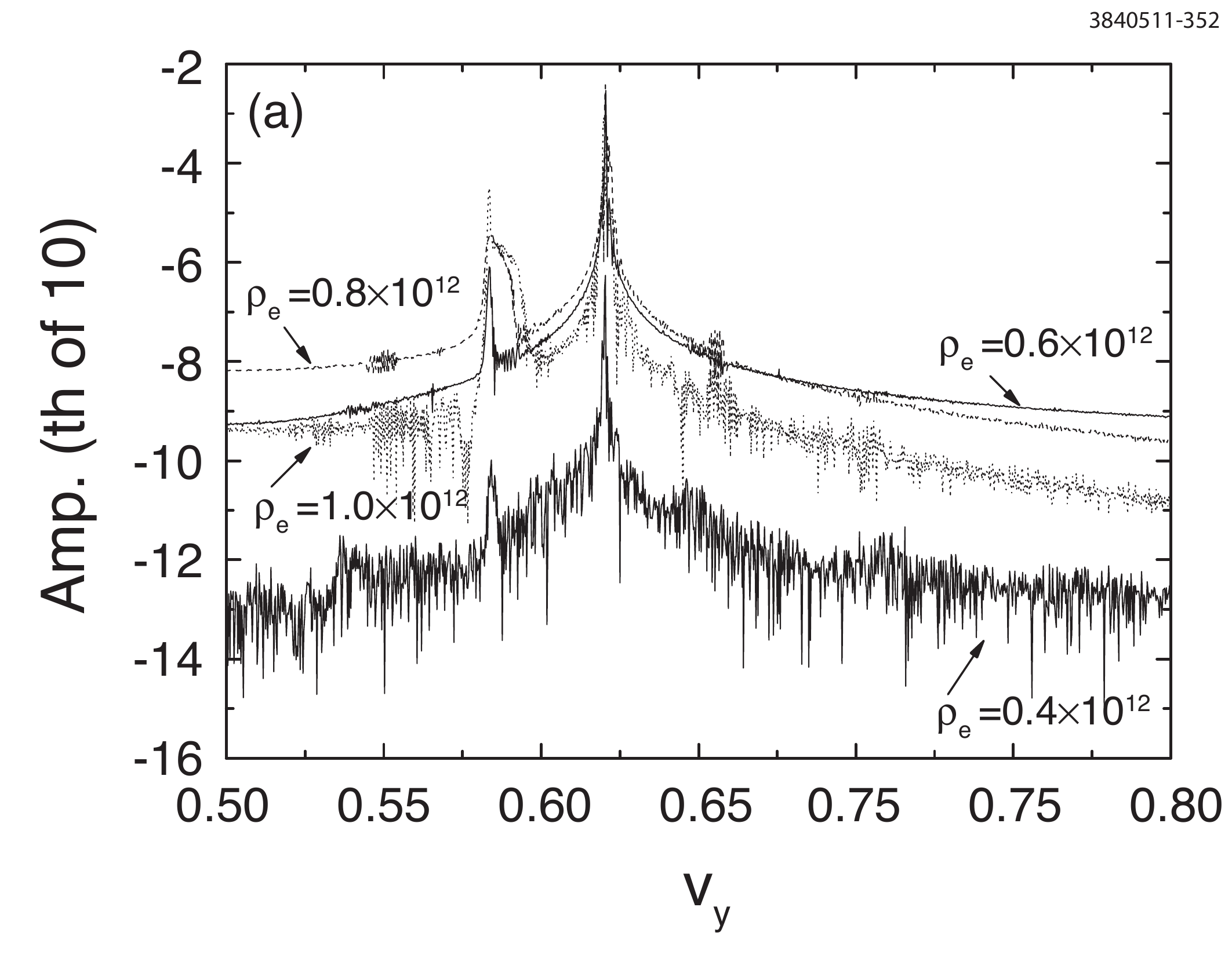} 
   \includegraphics[width=3.5in]{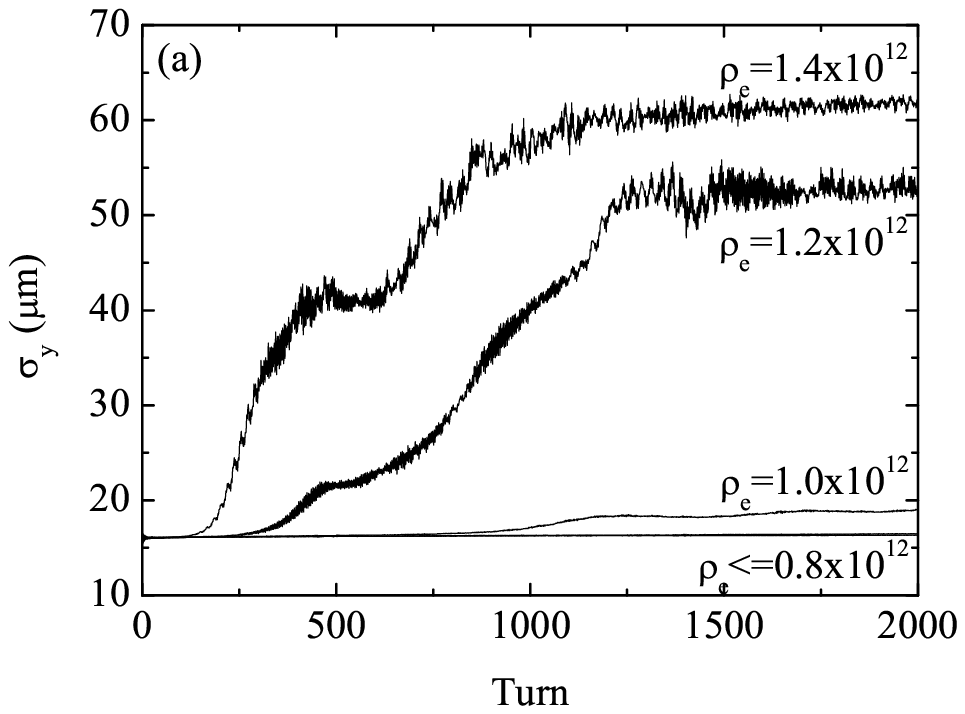}
   \end{array}$
   \caption{ \label{fig:ctaI8FFT} Left: Frequency spectra for the dipole moments for 2 GeV beam, simple lattice.
Vertical axis indicates the amplitude of frequency
spectra and the index corresponds to the power of 10. Right: Evolution of the beam size at  2  GeV, using realistic lattices. The cloud density $\rho_e$ is in units of m$^{-3}$. Both plots are PEHTS simulations from~\cite{Jin}.}
   \end{figure*}

  \begin{figure*}[ht!]
  \centering$
\begin{array}{cc}
  \includegraphics[width=3.3in]{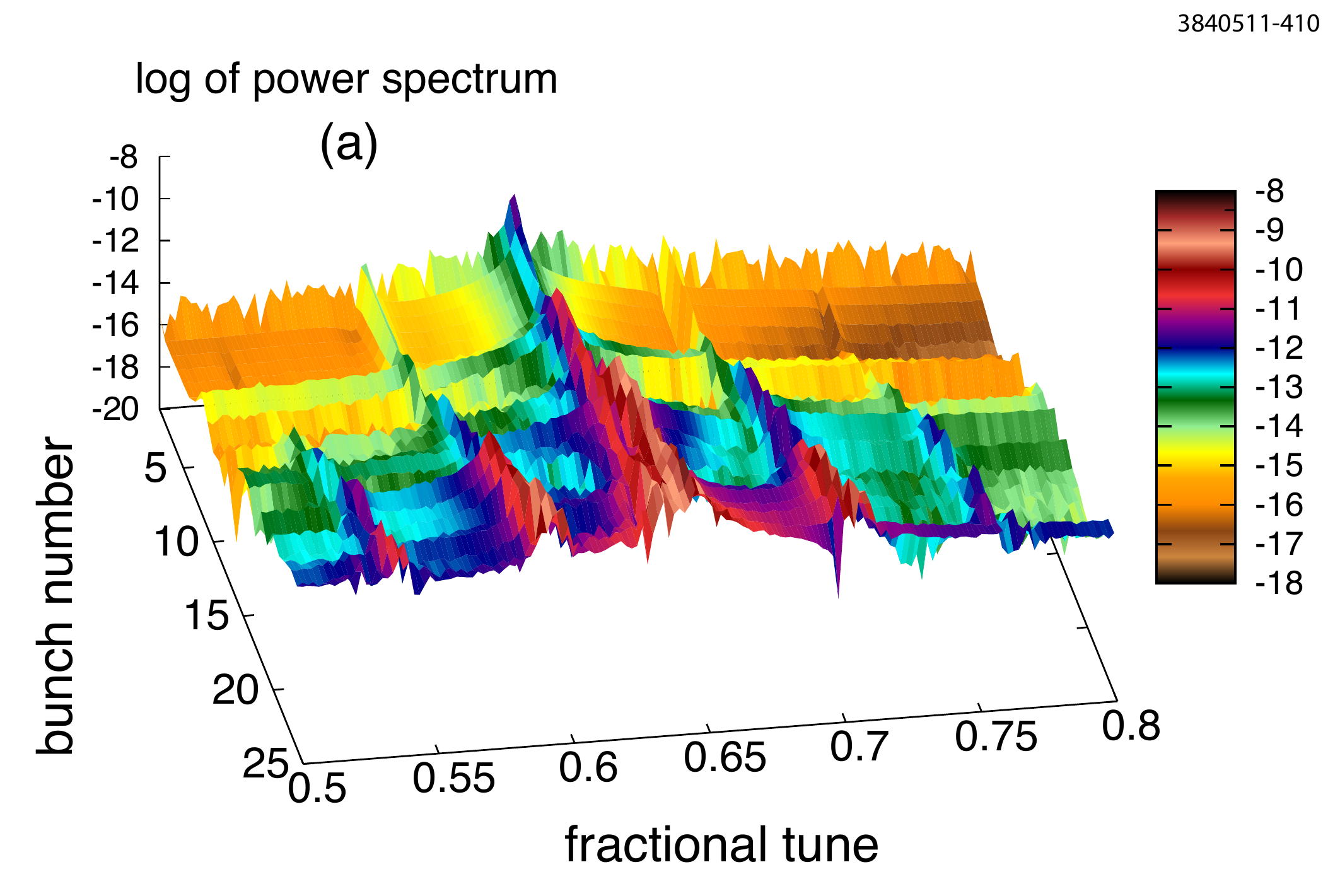} 
   \includegraphics[width=3.3in]{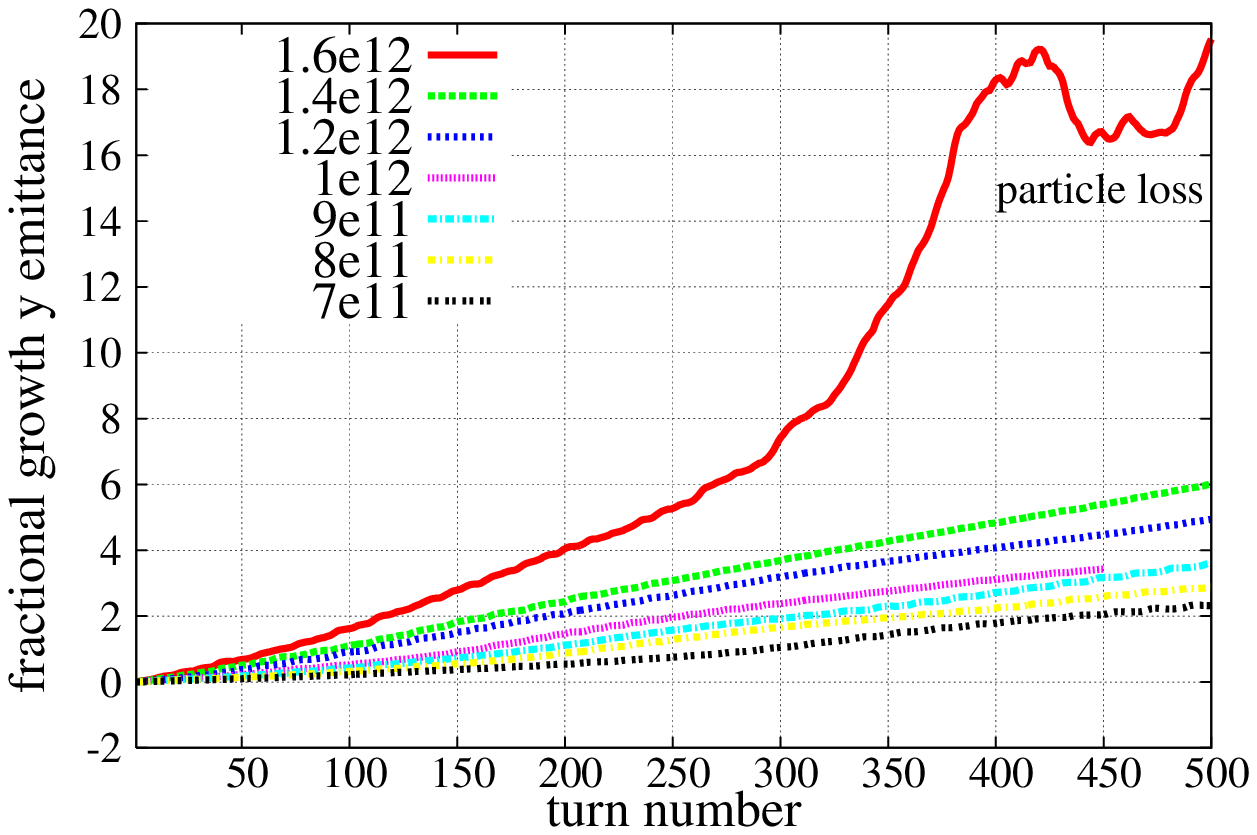}
   \end{array}$
   \caption{ \label{fig:ctaICMAD} Left: Frequency spectra for the dipole moments for 2 GeV beam, as function of bunch number.
Right: Evolution of the vertical emittance at  2  GeV. The cloud density $\rho_e$ is in units of m$^{-3}$. Both plots are CMAD simulations.}
   \end{figure*}

\subsection{Analytic estimates}
The analytic theory discussed in~\cite{Jin} was used to estimate the head-tail instability threshold in the coasting beam approximation. In this model, the electron cloud density corresponding to the instability threshold is given by

\begin{equation}
\label{eq:EC_BeamDynamics:anaest:e12}
\rho_{e,th}=\frac{2\gamma \nu_s \chi}{\sqrt{3} KQ r_e 
\left<\beta_{y}\right>C}.
\end{equation}
in which $\nu_s$ is the synchrotron tune, $K$ is the enhancement factor of the cloud density due to the ``pinch'', and $Q$ is the finite quality factor associated with the spread in electron oscillation frequencies. $\chi$ is the phase angle
of the electron oscillation in the field of the bunch:
\begin{equation}
\label{eq:EC_BeamDynamics:anaest:e8}
\chi=\frac{\omega_{e,y}\sigma_{z}}{c},
\end{equation}
where $\sigma_z$ is the bunch length, and $\omega_{e,y}$ is the frequency of electron motion within the field of the bunch, which depends on the bunch intensity and transverse size.

Guided by numerical simulations, 
$K$ is approximated as $\chi$, and $Q$ is taken to be the smaller of $Q_{nl}$ and $\chi$, in which the nonlinear resonator quality factor is estimated to be $Q_{nl}\sim 7$.

With these assumptions, the threshold of the electron cloud density for the instability becomes
\begin{equation}
\label{eq:EC_BeamDynamics:anaest:e15}
\rho_{e,th}=\frac{2\gamma \nu_s}{\sqrt{3} \mbox{Min}(Q_{nl},\chi) r_e \left<\beta_{y}\right>C}.
\end{equation}
Table~\ref{tab:thr} gives the key instability parameters for {\cesrta} at 2.1 and 4~GeV. At 2.1~GeV, the analytical estimate of the threshold density, $1.3\times10^{12}$~m$^{-3}$, is about 60\% higher than the measured threshold of $8\times10^{11}$~m$^{-3}$. At 4~GeV, the analytical estimate of the threshold density is $2.65\times10^{12}$~m$^{-3}$, about 30\% higher than the measured threshold of $2\times10^{12}$~m$^{-3}$.

\subsection{Numerical simulations}
\subsubsection{Coherent emittance growth}
Numerical simulations using PEHTS~\cite{Ohmi} have been done to refine the estimates of the threshold density at  2 GeV. These simulations~\cite{Jin} show both vertical emittance growth, and the presence of head-tail lines in the beam's dipole motion spectrum, above the threshold density (see Fig.~\ref{fig:ctaI8FFT}, left). The simulations show that dipole feedback is not able to suppress the emittance growth. The effects of dispersion, and a realistic lattice with 83 beam-cloud interaction points, were also studied. The threshold densities found for the realistic lattice (about $1.2\times10^{12}$~m$^{-3}$, see Fig.~\ref{fig:ctaI8FFT}, right)
were about 50\% higher than the analytical estimates\footnote{The beam parameters used for the simulations shown in Fig.~\ref{fig:ctaI8FFT} were different than those presented in Table~\ref{tab:thr}. For the parameters used for the simulations, the analytical estimates were about $0.8\times10^{12}$~m$^{-3}$.}.

Numerical simulations using CMAD~\cite{CMAD-ref, Kiran} were also done for 2 GeV beam energy. These simulations use a realistic lattice, with beam-cloud interaction points at every lattice element. As with PEHTS, they show both vertical emittance growth, and the presence of head-tail lines in the beam's dipole motion spectrum (see Fig.~\ref{fig:ctaICMAD}, left). For the same cloud density above the head-tail threshold, CMAD and PEHTS predict the same level of vertical emittance growth after 500 turns, within a factor of 2 (see Fig.~\ref{fig:ctaICMAD}, right).

\subsubsection{Incoherent emittance growth}

For the realistic lattice, PEHTS was also used to estimate incoherent emittance growth below the head-tail threshold. At the
electron-cloud density  $0.8\times 10^{12}$~m$^{-3}$ for 2 GeV beam energy,
the growth rate of the beam size is about $1.9\times 10^{-4}$ $\mu$m/turn. This
is smaller than the radiation damping rate in {\cesrta} at 2~GeV, $4.6\times
10^{-5}\sigma_{y}$/turn. Assuming the electron cloud growth rate is independent of beam size, the equilibrium vertical beam size will increase by a factor of
$\sim1.25$ due to sub-threshold incoherent emittance growth.
Beam size increases of this magnitude should be observable at {\cesrta}. For example, the slow growth along the train seen in Fig.~\ref{fig:welh08_fig235} (top), at a bunch current below the coherent instability threshold, may be due to incoherent emittance growth. Experimental studies to look for such emittance growth are planned for the future.

\begin{figure*}[htb!]
  \centering
  \includegraphics[width=6.6in]{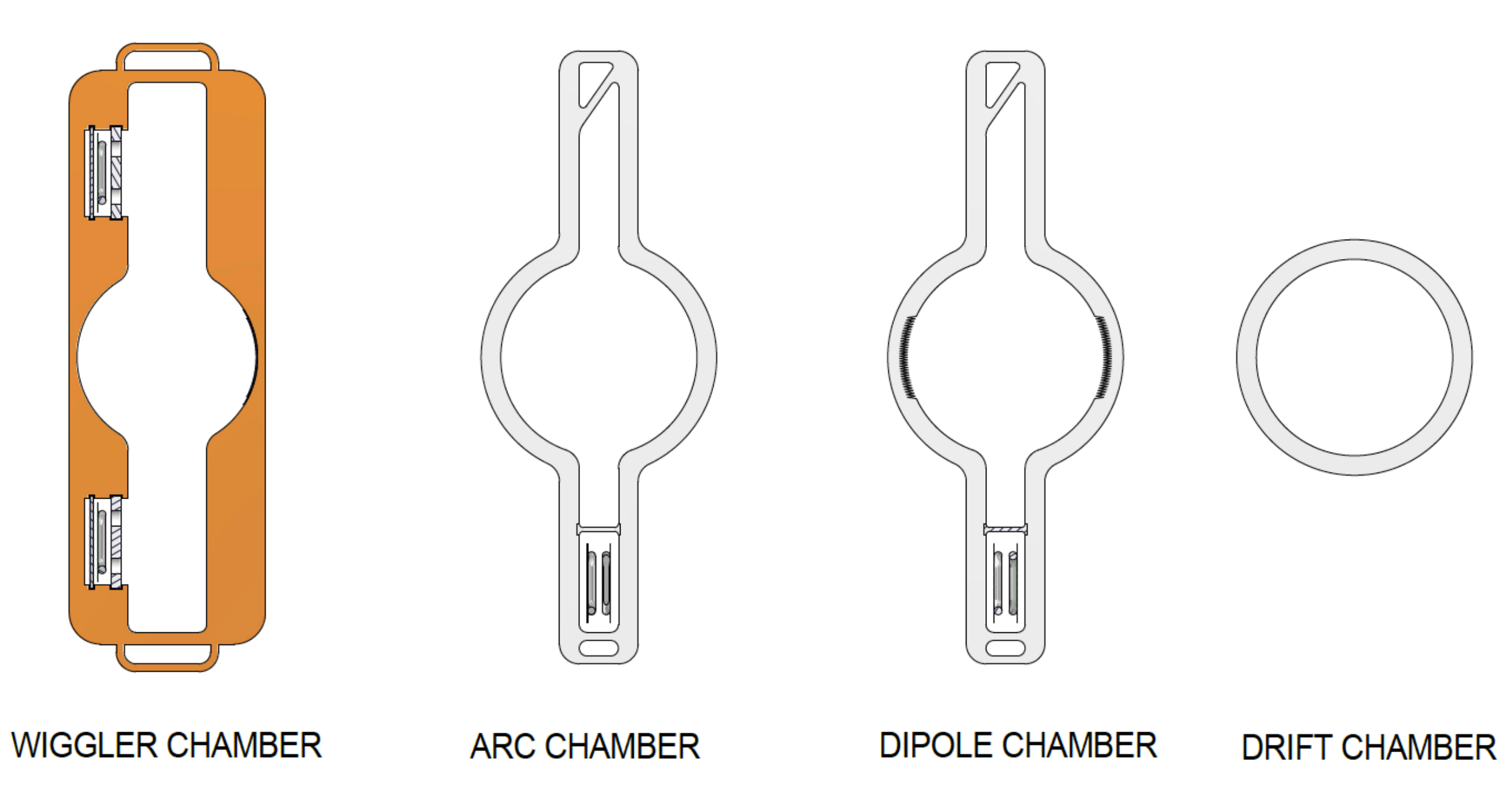} 
   \caption{ \label{fig:ILCDR_VC} Vacuum chamber designs for magnetic elements in the ILC damping ring. The upper part of the figures correspond to the outer radius of the ring, when installed.}
   \end{figure*}

\section{Outlook for the ILC}

In Table~\ref{tab:thr}, the third column shows the parameters of the ILC positron damping ring, and provides an estimate of the threshold for the electron-cloud-induced head-tail instability, using the simple analytical formula which was shown to give approximate agreement with measurements at {\cesrta}. The estimated threshold, $2.3\times 10^{11}$ m$^{-3}$, is quite low, primarily because of the low momentum compaction. Thus, rigorous suppression of the electron cloud will be required to achieve the design emittance in this ring. This suppression is accomplished by
\begin{itemize}
\item Suppressing the primary photoelectrons which seed the cloud, through the use of antechambers and photon stops; and
\item Suppressing the secondary electrons which build up the cloud near the beam, through the use of coatings with a low secondary emission yield, solenoids, grooves, and clearing electrodes.
\end{itemize}
The overall suppression is designed to reduce the electron cloud density near the beam to a level which is well below the instability threshold.

\subsection{Photoelectron suppression}
The recently developed code \texttt{Synrad3D}~\cite{S3D,S3D2} was used to estimate the distribution of synchrotron radiation photon absorption sites around the ring, which determines the initial distribution of photoelectrons.   The design vacuum chamber geometry, including all feature details such as antechambers and photon stops, is used in the calculation. Both specular and diffuse scattering are included in the simulation. For the scattering calculation, the surface material is approximated as an aluminum surface with a thin carbon coating\footnote{The presence of this coating is suggested by X-ray scattering data from a conditioned technical vacuum chamber surface.}, and the surface roughness parameters that were assumed (rms roughness 100 nm, transverse correlation length 5000 nm) are representative of a typical technical vacuum chamber. For these parameter, diffuse scattering dominates over specular reflection.

The vacuum chamber design~\cite{Conway} is shown in Fig.~\ref{fig:ILCDR_VC}. The antechambers shown in this figure are designed to trap most of the synchrotron radiation photons, thereby suppressing the photons absorbed in the chamber itself.  To achieve the required level of suppression in the wiggler regions, it was necessary to use an antechamber with a 20 mm gap. In the arc and dipole chambers, a 10 mm gap was sufficient, but  a 45$^{\circ}$ slope had to be introduced into the antechamber back walls, as shown in Fig.~\ref{fig:ILCDR_VC}, to inhibit scattering out of the antechamber.
  
The photon intensity distributions predicted by \texttt{Synrad3D} for
magnetic elements in one of the arcs of the damping ring are shown in Fig.~\ref{fig:photons}. The absence of photons at zero and $\pi$ radians are due to the
antechambers. The top-bottom asymmetry is due to the antechamber back wall slope.
  \subsection{Secondary electron suppression}
The vacuum chamber incorporates mitigation techniques~\cite{Pivi} in each magnetic field
region to suppress the buildup of the electron cloud in
the vacuum chambers of the ring. The mitigation methods
were selected based on the results of an intense experimental research
effort at {\cesrta}, KEKB and PEP-II. In the arc regions of the ring, the
50-mm aperture vacuum chambers utilize TiN-coating to
suppress the secondary electron yield of the chambers. In the dipoles, the
electron cloud is further suppressed by the use of longitudinal grooves
on the top and bottom surfaces. In the wiggler region,
a 46-mm aperture chamber utilizes clearing electrodes to suppress growth of the cloud. In the straights of the ring, the 50-mm aperture round vacuum chamber is TiN-coated
for secondary suppression. Drift regions throughout
the ring will employ solenoid windings to further suppress
cloud buildup near the beam.   

\subsection{Estimates of the ringwide electron cloud density}
\begin{figure}[htb!]
   \centering
    \includegraphics[width=3.2in]{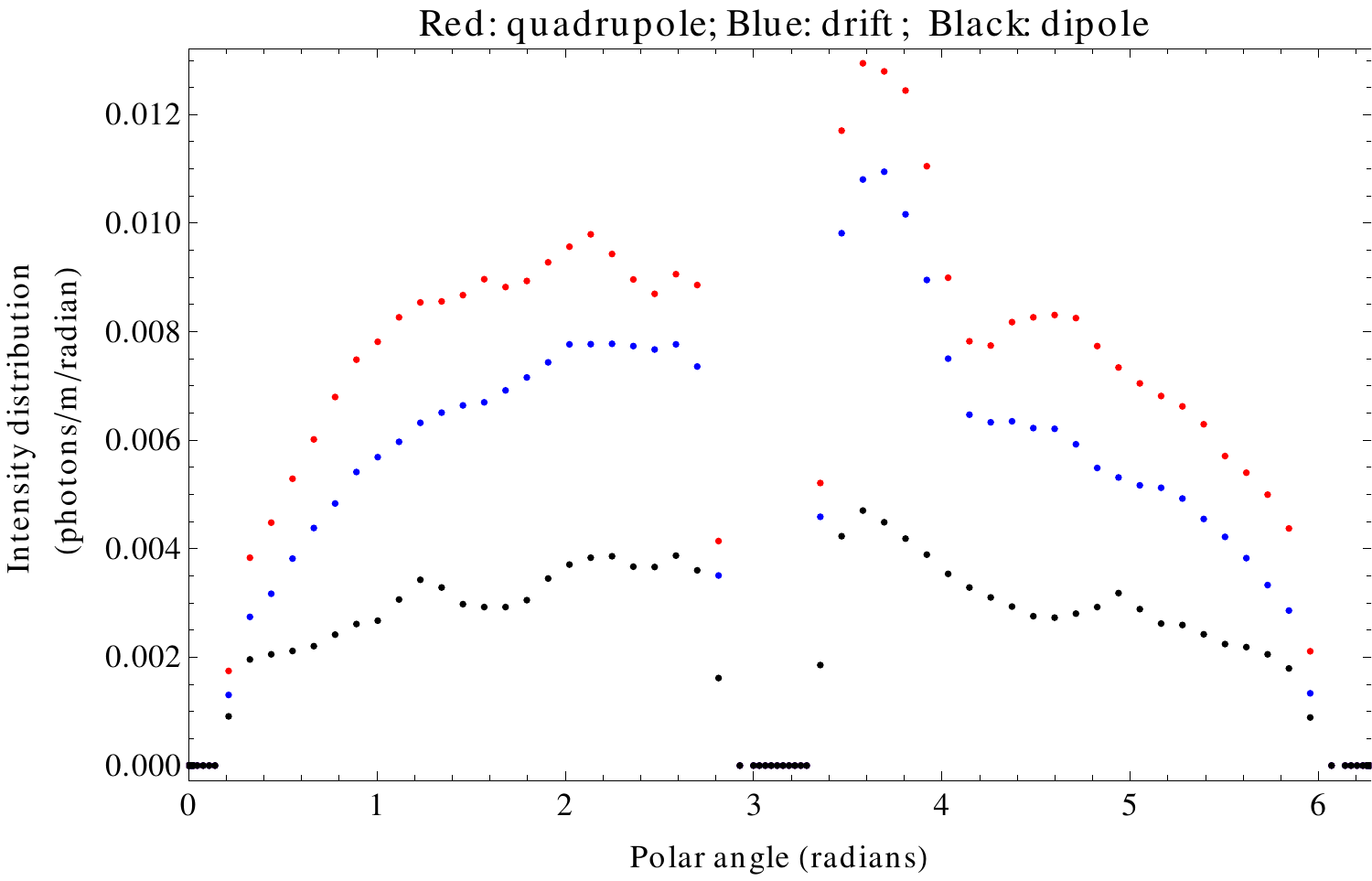}
   \caption{\label{fig:photons}Intensity distributions for absorbed photons in
one arc of the damping ring, averaged over three types of
magnetic environments. The average over the quadrupole
regions is shown in red. The average over the field-free
regions is shown in blue and the average over the dipole
regions is shown in black. The angle is defined to be zero
where the vacuum chamber intersects the bend plane on the
outside of the ring. The angle $\pi/2$ corresponds to the top
of the vacuum chamber.}
\end{figure}

Using the photon distributions shown in Fig.~\ref{fig:photons}, the electron cloud buildup codes \texttt{POSINST}, \texttt{ECLOUD} and \texttt{CLOUDLAND} have been used to estimate the equilibrium electron cloud density buildup in different magnetic environments in the ILC damping ring. In each environment, the appropriate secondary electron mitigation technique, as specified in Reference~\cite{Pivi}, has been assumed. The detailed results are presented in Reference~\cite{IPAC12}.

In the 91 m of wigglers, the clearing electrodes effectively reduce the cloud density to a negligible level. Similarly, the TiN coating and the solenoids in the drift regions (total length $\sim$2200 m) also suppress the cloud density near the beam in these regions quite effectively. The average cloud density within $20\sigma$ of the beam in the $\sim$450 m
of dipoles is $0.4\times10^{11}$ m$^{-3}$ (assuming no grooves). In the $\sim$450 m of arc and straight section quadrupoles and sextupoles, the average cloud density within $20\sigma$ of the beam is about $1.5\times10^{11}$ m$^{-3}$. The density in the quadrupoles located in the wiggler regions is quite high $(\sim12\times10^{11}$ m$^{-3}$) but their length is only 18 m. 

Overall, the ringwide average cloud density near the beam is about $0.35\times10^{11}$ m$^{-3}$. This estimate is based on simulations of dipole chambers without grooves, and so it is an overestimate of the density. If the grooves reduced the cloud density in the dipoles to zero, the ringside average would fall to below $0.3\times10^{11}$ m$^{-3}$.
\subsection{Comparison with the head-tail instability threshold}
From Table~\ref{tab:thr}, the head-tail instability threshold is $2.3\times 10^{11}$ m$^{-3}$. For {\cesrta}, this simple formula overestimates the threshold by up to 60\%. Thus, the threshold may be as low as $\sim1.5\times 10^{11}$ m$^{-3}$. This is still at least 4 times larger than the estimated cloud density in the ring, if the photon suppression and secondary electron mitigations perform as simulated. Thus, it appears that the ILC damping ring should not suffer from an electron-cloud induced coherent instability. There may still be emittance growth due to incoherent effects, however. It may be limited since the cloud density is quite low. This is a subject that requires additional study.

%%%%%%%%%%%%%%%
\section{Summary and Conclusions}
\begin{itemize}
\item The {\cesrta} research program has investigated the dynamics of trains of positron bunches in the presence of the electron cloud through measurements of bunch-by-bunch coherent tune shifts, frequency spectra, and beam size. 
\item Coherent tune shifts have been compared with the predictions of cloud buildup models (augmented with a new code, \texttt{Synrad3D}, to characterize the photoelectrons) in order to validate the buildup models and determine their parameters.
\item Frequency spectra have been used to determine the threshold density above which signals for electron-cloud-induced head-tail instabilities develop.
\item An X-ray beam size monitor has been used to determine the conditions under which beam size growth occurs, and to correlate these observations with the frequency spectral measurements.
\item Simulation codes have been used to model the cloud-induced head-tail instability. The predicted features of the instability agree reasonably well with the measurements.
\item The success of the cloud buildup and head-tail instability codes in modeling the observations gives confidence that these codes can be used to accurately predict the performance of future storage rings. 
\item For the ILC damping ring, an antechamber design has been developed with the help of the \texttt{Synrad3D} photon simulation. 
\item EC buildup simulations have been done based on the ILC damping ring EC mitigation prescriptions.
\item The expected ringwide average cloud density is around $0.35\times10^{11}$ m$^{-3}$, dominated by cloud in the quadrupoles and sextupoles.
\item Based on simple analytic estimates and extrapolation from {\cesrta}, the instability threshold in the ILC damping ring should be 4-5 times higher than the expected ringwide average cloud density.
\item Incoherent emittance growth for the ILC damping ring has yet to be fully studied, but is expected to be small because of the low ringwide average cloud density.
\end{itemize}

%                \begin{thebibliography}{9}   % Use for  1-9  references

\end{document}